
\input phyzzx.tex
\pubnum{TIFR/TH/94-15}
\date{12 April, 1994}
\titlepage
\title{\bf NUMERICAL DIAGONALIZATION STUDY OF THE TRIMER
DEPOSITION-EVAPORATION MODEL IN ONE DIMENSION}
\author{Peter B. Thomas\foot{Present address: Department of Physics,
Brookhaven National Laboratory, Upton, \break NY 11973; Internet:
thomas@cmth.phy.bnl.gov},
M. K. Hari Menon\foot{Internet address: hari@theory.tifr.res.in} and
Deepak Dhar\foot{Internet address: ddhar@theory.tifr.res.in}}
\address {Theoretical Physics Group, \break
Tata Institute of Fundamental Research, \break
Homi Bhabha Road, Bombay 400 005, India.}
\abstract

We study the model of deposition-evaporation of trimers on a line
recently introduced by Barma, Grynberg and Stinchcombe. The stochastic
matrix of the model can be written in the form of the Hamiltonian of a
quantum spin-$1\over 2$ chain with three-spin couplings given by $ H=
\sum\displaylimits_i ~[~(1-\sigma_i^-\sigma_{i+1}^-\sigma_{i+2}^-)~
\sigma_i^+\sigma_{i+1}^+\sigma_{i+2}^+ + h.c~] $.
We study by exact numerical diagonalization of $H$ the variation of
the gap in the eigenvalue spectrum with the system size for rings of
size up to $30$. For the sector corresponding to the initial condition
in which all sites are empty, we find that the gap vanishes as
$L^{-z}$ where the gap exponent $z$ is approximately
$2.55\pm 0.15$. This model is equivalent to an interfacial roughening
model where the dynamical variables at each site are matrices. From our
estimate for the gap exponent we conclude that the model belongs to a new
universality class, distinct from that studied by Kardar, Parisi and
Zhang.

DAE\vskip .4cm
\noindent PACS numbers: 02.50+s, 75.10.J, 82.20.M, 05.50+q

\endpage

\Ref\mba{M. Barma, M. D. Grynberg and R. B. Stinchcombe, {\it Phys. Rev.
Lett.} {\bf 70}, 1033 (1993).}

\Ref\mbb{R. B. Stinchcombe, M. D. Grynberg and M. Barma, {\it Phys. Rev.
E}, {\bf 47} 4018 (1993)}

\Ref\ddmba{D. Dhar and M. Barma {\it Pramana} {\bf 41} L193 (1993).}

\Ref\ddmbb{ M. Barma and D. Dhar {\it Proceedings of the  Solid State
Physics Symposium} {\bf 36C} 489 (1993).}

\Ref\prica{ M. Barma and D. Dhar , {\it In preparation}}

\Ref\kpz{M. Kardar, G. Parisi and Y. C. Zhang {\it Phys. Rev. Lett.}
{\bf 56} 889 (1986).}

\Ref\doh{J. P. Doherty, M. A. Moore, J. M. Kim and A. J. Bray
{\it Manchester Preprint} (1993).}

Many physical processes such as heterogeneous catalysis, chemical
reactions on polymer chains, adsorption on solid surfaces, {\it etc.}
involve evaporation and deposition of reactants on a substrate.
Recently Barma {\it et al. } have introduced a simple model which
shows that the excluded volume effect together with dissociation and
re-combination of the reactants on the surface can give rise to very
interesting dynamical behaviour. In their model they have studied a
random deposition-evaporation process of $k$ identical atoms (called
$k$-mers, $k=1,2,3...$) on a surface [\mba,\mbb].  It has been shown
that in one-dimension when $k\ge3$ the phase space breaks up into an
exponentially large number of dynamically disconnected sectors and the
model has infinite number of conserved quantities. It is found that in
this case the auto-correlation function in the steady state decays
with time $t$ as $t^{-1/4}$, $t^{-1/2}$, $t^{-0.59}$ or as
$e^{-\sqrt{t}}$, depending on the initial condition.  The behaviour of
the auto-correlation function for different initial conditions is
understood in terms of the random walk of the substrings which
constitutes what is called an irreducible string [\ddmbb]. However for
the steady state corresponding to the empty configuration as the
initial condition, this analysis does not apply. In this case, for
trimer model, Monte Carlo simulations show power law decay of
autocorrelation function with an approximate value for the exponent
$0.59$ [\prica]. A theoretical understanding of this exponent is still
lacking. Thus our main motivation is to understand the dynamics of
trimer model in this sector. We have done a study of the trimer model
on a one dimensional lattice, by exact diagonalisation of the
stochastic matrix in this sector.

In this letter we restrict ourselves to the study of trimers (k=3) on
a  line (d=1). We consider a ring of $L$ sites. At each site $i$ is
a dynamical variable $n_i$ which takes values $0$ or $1$, depending on
whether the site is occupied or not. In time a configuration $\{n_i\}$ evolves
stochastically by Markovian dynamics as follows: Any three adjacent
empty sites can become occupied with a rate $\epsilon$ and any three
adjacent occupied site can become empty with a rate $\epsilon'$.

If $P(C,t)$ is the probability that the ring has configuration $C$ at
time $t$, then $P(C,t)$ satisfies the master equation
$$
{\partial \over \partial t} P(C,t) = \sum_{C'} W_{CC'} P(C',t)
\eqn\zeroa\relax
$$
Where the transition  rate matrix $\hat W$ for the case $\epsilon
=\epsilon'$ can be written as
$$
 \hat W = \epsilon\,\sum_{i=1}^L~ \left[(1-\sigma_i^-\sigma_{i+1}^-
\sigma_{i+2}^-)~\sigma_i^+\sigma_{i+1}^+\sigma_{i+2}^+ + \hbox{h.c.}\right]
\eqn\zeroc\relax
$$
where $\sigma_i^-$ and $\sigma_i^+$ are the Pauli annihilation and
creation operators at site $i$.

Since $\hat W$ is a stochastic matrix where the transition rates
satisfy detailed balance, all its eigenvalues are real and non-
positive.  The infinite number of conservation laws of this
Hamiltonian can be encoded into a single conservation law of the
irreducible string [\ddmba].  For any configuration the irreducible string is
defined as follows: From the L-bit string of $0$'s and $1$'s
representing the configuration, we recursively delete any consecutive
occurrence of three $0$'s or $1$'s until no further deletions are
possible. The irreducible string is conserved under dynamics and can
be used to label uniquely each of the dynamically disconnected
sectors.  There is a large degeneracy for the eigenvalue $0$,
reflecting the large number of conservation laws in the model. An
example of an eigenvector with zero eigenvalue is any a configuration
which has no $3$ adjacent $0$'s or $1$'s. Such a state cannot evolve in
time. The number of such configurations has been shown to vary as
$\mu^L$ for large $L$, where $\mu$ is the golden mean $(\sqrt{5}+1)/2$
[\ddmba].

We can exactly diagonalise $\hat W$ in some almost totally
jammed sectors. For example, if the sector corresponds to an
irreducible string of length $L-3$, then it is easy to see that the
corresponding stochastic matrix in general has size of $O(L^2)$. Under
dynamics the position of the reducible block on the ring changes and
its motion can be described as a random walk. In this case it can be
shown that the mean square displacement increases linearly with time.
This corresponds to a dynamical exponent of $z=2$.  Sectors with
irreducible string length $L-6$ correspond to diffusion of $2$
interacting random walkers. In this case the size of the stochastic
matrix will be of $O(L^3)$. When the two walkers are next to each
other, they stay there longer, which corresponds to an attractive
interaction. The dynamical exponent will be $2$ in this case also.

The most interesting sector corresponds to the case when the length of
the irreducible string $(l)$ is very small compared to $L$. In this case
Monte Carlo simulations [\prica] have shown that the attractive
interaction between these ``random walkers'' gives rise to a
sub-diffusive behavior, with the dynamical exponent $z>2$. In this
paper, we estimate this exponent by numerically diagonalising the
stochastic matrix for small systems and assuming finite size scaling.

For numerical diagonalisation it is desirable to reduce the size of
the matrix as much as possible by making use of the known symmetries
and conservation laws of the model. For periodic boundary conditions,
and for the special case of deposition and evaporation rates equal
($\epsilon=\epsilon'$), in addition to the conservation law of the
irreducible string, one can make use of the three symmetries of
the system namely translation, reflection and flip, to reduce the size
of the matrix by about a factor of $2L$. Let $\hat T$, $\hat P$ and
$\hat F$ be the operators corresponding to these symmetries. They are
defined by
\def\s#1{\scriptscriptstyle {#1}}
\def\conf{n_{\s 1},n_{\s 2},..,n_{\s i},..,n_{\s L}}
$$ \eqalign{
\hat T \ket\conf &= \ket{n_{\s 2},n_{\s 3},..,
       n_{\s {i+1}},..,n_{\s L},n_{\s 1}} \cr
\hat P \ket\conf &= \ket{n_{\s L},n_{\s {L-1}},..,n_{\s i},..,n_{\s 1}} \cr
\hat F \ket\conf &= \ket{\bar n_{\s 1},\bar n_{\s 2},..,\bar n_{\s i},..,
       \bar n_{\s L}}; \qquad \hbox{where } \bar n_{\s i} = 1-n_{\s i} \cr}
\eqn\two\relax
$$
Here $\ket\conf$ is a vector in the Hilbert space representing the
configuration $\{n_i\}$. These operators satisfy the following algebra
$$ \eqalign{
&[\hat T,\hat F] = [\hat P,\hat F]=0 \cr
&\hat T^L = \hat P^2 = \hat  F^2 = 1 \cr
&\hat T \hat P = \hat P \hat T^{-1} \cr}
\eqn\three\relax
$$
Note that $\hat T$ and $\hat P$ do not commute. The three operators
which simultaneously commute with $\hat W$ and with each other are
$\hat F$, $\hat P$ and $\left(\hat T + \hat T^{-1}\right)$. Let their
corresponding eigenvalues be $f$, $p$ and $2\cos(k)$ respectively,
where $f=\pm 1$, $p=\pm 1$ and $k=2n\pi/L$; $n=0,1,...,L-1$. The
simultaneous eigenvectors of these three operators are of the form
$$
\eqalign{
\ket{k,f,p,+} &= (1+f\hat F)(1+p\hat P)\sum_{r=1}^L T^{r} \cos(kr)
\ket{C} \cr
\ket{k,f,p,-} &= (1+f\hat F)(1+p\hat P)\sum_{r=1}^L T^{r} \sin(kr)
\ket{C} \cr}
\eqn\five\relax
$$ where $\ket{C}$ is any of the vectors $\ket{\{n_i\}}$.

We have used the states \five\ as the basis for the stochastic
matrix.  For the null sector, the matrix splits into $2L$ blocks,
corresponding to combinations of the $2$ eigenvalues of $\hat F$ and
the $L$ eigenvalues of $\hat T$.  Of these, due to a Kramers type
degeneracy in the eigenvalues for the momentum values $k$ and
$2\pi-k$, we can fix $p$ to always be equal to unity, and sweep over
only half of the allowed momentum values. For lattice lengths which
are not multiples of three, there is an additional degeneracy in the
eigenvalues for $f=1$ and $f=-1$, since these states and their flipped
counterparts are not connected by the dynamics. Since the size of the
null sector $\sim (27/4)^{L/3}L^{-3/2}$ [\ddmba], the size of each block
$\sim (27/4)^{L/3}L^{-5/2}$. For any lattice length, each block of
the matrix is real and sparse, since all rows or columns  have
at most $L$ non-zero entries.

The difference between the largest and the second largest eigenvalue
of the complete matrix is proportional to the inverse relaxation time.
The largest eigenvalue is zero and it lies in the block $k=0$, $f=1$.
To find the second largest eigenvalue of the full matrix, we have
numerically computed the largest eigenvalue in all the other blocks,
and the second-largest eigenvalue in the $k=0$, $f=1$ block.
Simple iteration of the eigenvector after suitably shifting all the
eigenvalues, converged sufficiently fast for these blocks.  This
method preserves the sparseness of the blocks, which is necessary to
keep the memory requirement of the program as low as possible.  For
the $k=0$, $f=1$ block, we computed the second largest eigenvalue, by
ensuring orthogonality of the iterated vector to the eigenvector
corresponding to the zero eigenvalue.

We have computed these eigenvalues for lattice sizes ranging from
$L=3$ to $L=30$. When $L$ is a multiple of $3$, the irreducible string
in the null sector has length zero and in this case we have
diagonalized the stochastic matrix for both the $f=1$ and $f=-1$
case. For the case $f=-1$ the smallest eigenvalue occurs for $k=2\pi(1
- 1/L)/3$, and for the case $f=1$ it occurs for $k=2\pi/3$.  When
$L=3n+1$ and $L=3n+2$, where $n$ is an integer, the irreducible string
in the sector where the initial state is all empty has length $1$ and
$2$ respectively.  In this case, as explained earlier, the eigenvalues
for $f=1$ and $f=-1$ are degenerate. We have estimated the gap exponent
$z$ for each of these $4$ sets of data, by assuming the scaling
relation $\lambda \sim L^{-z}$.  We define the effective exponent
$$
z_{\scriptscriptstyle L} ={\log[\lambda_{\hbox{\tenrm L}-3}
/\lambda_{\hbox{\tenrm L}}]
\over {\log[L/(L-3)]}}.
\eqn\six\relax
$$
The sizes of the matrices, eigenvalues and estimate of the dynamical
exponent $z_{\scriptscriptstyle L}$ are shown in tables below. The
$z_{\scriptscriptstyle L}$ values are also plotted as a function of $1/L$
in figure 1.


\vskip30pt\hbox{
{\vbox{\tabskip=0pt \offinterlineskip
\halign to 100pt{\strut#& \vrule# \tabskip=1em plus 2em&
   \hfil#& \vrule#\tabskip=0pt\cr\noalign{\hrule}

&&\omit\hidewidth Length of \hidewidth& \cr \noalign{\hrule width0.4pt}
&&\omit\hidewidth the lattice \hidewidth& \cr\noalign{\hrule}

&&  3 & \cr\noalign{\hrule}
&&  6 & \cr\noalign{\hrule}
&&  9 & \cr\noalign{\hrule}
&& 12 & \cr\noalign{\hrule}
&& 15 & \cr\noalign{\hrule}
&& 18 & \cr\noalign{\hrule}
&& 21 & \cr\noalign{\hrule}
&& 24 & \cr\noalign{\hrule}
&& 27 & \cr\noalign{\hrule}
&& 30 & \cr\noalign{\hrule}
}}}
\hskip-4.4pt
{\vbox{\tabskip=0pt \offinterlineskip
\halign to 300pt{\strut#& \vrule# \tabskip=1em plus 2em&
   \hfil#& \vrule#&
   \hfil#& \vrule#&
   \hfil#& \vrule# \tabskip=0pt\cr\noalign{\hrule}

&&\multispan5\hfil $f=-1$ \hfil
& \cr\noalign{\hrule}
&&\omit\hidewidth  Matrix Size \hidewidth&&
  \omit\hidewidth $\lambda_{\hbox{\tenrm min}}$ \hidewidth&&
  \omit\hidewidth $z_{\scriptscriptstyle L}$ \hidewidth& \cr\noalign{\hrule}

&& $1     $ && $-6.00000$ && $       $ & \cr\noalign{\hrule}
&& $1     $ && $-2.00000$ && $1.58492$ & \cr\noalign{\hrule}
&& $2     $ && $-0.87113$ && $2.04977$ & \cr\noalign{\hrule}
&& $10    $ && $-0.43876$ && $2.38400$ & \cr\noalign{\hrule}
&& $35    $ && $-0.26065$ && $2.33375$ & \cr\noalign{\hrule}
&& $170   $ && $-0.16932$ && $2.36607$ & \cr\noalign{\hrule}
&& $815   $ && $-0.11744$ && $2.37373$ & \cr\noalign{\hrule}
&& $4176  $ && $-0.08545$ && $2.37929$ & \cr\noalign{\hrule}
&& $21872 $ && $-0.06455$ && $2.38333$ & \cr\noalign{\hrule}
&& $118175$ && $-0.05020$ && $2.38672$ & \cr\noalign{\hrule}
}}}
}


pt\vskip30pt\hbox{
{\vbox{\tabskip=0pt \offinterlineskip
\halign to 100pt{\strut#& \vrule# \tabskip=1em plus 2em&
   \hfil#& \vrule#\tabskip=0pt\cr\noalign{\hrule}

&&\omit\hidewidth Length of \hidewidth& \cr \noalign{\hrule width0.4pt}
&&\omit\hidewidth the lattice \hidewidth& \cr\noalign{\hrule}

&&  6 & \cr\noalign{\hrule}
&&  9 & \cr\noalign{\hrule}
&& 12 & \cr\noalign{\hrule}
&& 15 & \cr\noalign{\hrule}
&& 18 & \cr\noalign{\hrule}
&& 21 & \cr\noalign{\hrule}
&& 24 & \cr\noalign{\hrule}
&& 27 & \cr\noalign{\hrule}
&& 30 & \cr\noalign{\hrule}
}}}
\hskip-4.4pt
{\vbox{\tabskip=0pt \offinterlineskip
\halign to 300pt{\strut#& \vrule# \tabskip=1em plus 2em&
   \hfil#& \vrule#&
   \hfil#& \vrule#&
   \hfil#& \vrule# \tabskip=0pt\cr\noalign{\hrule}
&&\multispan5\hfil $f=1$ \hfil& \cr\noalign{\hrule}
&&\omit\hidewidth  Matrix Size \hidewidth&&
  \omit\hidewidth $\lambda_{\hbox{\tenrm min}}$ \hidewidth&&
  \omit\hidewidth $z_{\scriptscriptstyle L}$ \hidewidth& \cr\noalign{\hrule}

&& $1     $ && $-2.00000$ &&$       $& \cr\noalign{\hrule}
&& $2     $ && $-1.25553$ &&$1.14828$& \cr\noalign{\hrule}
&& $10    $ && $-0.67412$ &&$2.16180$& \cr\noalign{\hrule}
&& $35    $ && $-0.44217$ &&$1.88375$& \cr\noalign{\hrule}
&& $173   $ && $-0.29577$ &&$2.21307$& \cr\noalign{\hrule}
&& $811   $ && $-0.20803$ &&$2.28276$& \cr\noalign{\hrule}
&& $4186  $ && $-0.15213$ &&$2.34360$& \cr\noalign{\hrule}
&& $21874 $ && $-0.11485$ &&$2.38637$& \cr\noalign{\hrule}
&& $118175$ && $-0.08903$ &&$2.41725$& \cr\noalign{\hrule}
}}}
}


\vskip30pt\hbox{
{\vbox{\tabskip=0pt \offinterlineskip
\halign to 100pt{\strut#& \vrule# \tabskip=1em plus 2em&
   \hfil#& \vrule#\tabskip=0pt\cr\noalign{\hrule}

&&\omit\hidewidth Length of \hidewidth& \cr
&&\omit\hidewidth the lattice \hidewidth& \cr\noalign{\hrule}

&&  4 & \cr\noalign{\hrule}
&&  7 & \cr\noalign{\hrule}
&& 10 & \cr\noalign{\hrule}
&& 13 & \cr\noalign{\hrule}
&& 16 & \cr\noalign{\hrule}
&& 19 & \cr\noalign{\hrule}
&& 22 & \cr\noalign{\hrule}
&& 25 & \cr\noalign{\hrule}
}}}
\hskip-4.4pt
{\vbox{\tabskip=0pt \offinterlineskip
\halign to 300pt{\strut#& \vrule# \tabskip=1em plus 2em&
   \hfil#& \vrule#&
   \hfil#& \vrule#&
   \hfil#& \vrule# \tabskip=0pt\cr\noalign{\hrule}

&&\hfil&&\hfil&&\hfil&\cr
&&\omit\hidewidth Matrix Size \hidewidth&&
  \omit\hidewidth $\lambda_{\hbox{\tenrm min}}$ \hidewidth&&
  \omit\hidewidth $z_{\scriptscriptstyle L}$ \hidewidth & \cr\noalign{\hrule}

&& $1    $ && $-1.00000$ && $       $ & \cr\noalign{\hrule}
&& $4    $ && $-0.23008$ && $2.62557$ & \cr\noalign{\hrule}
&& $17   $ && $-0.09277$ && $2.54655$ & \cr\noalign{\hrule}
&& $84   $ && $-0.04754$ && $2.54625$ & \cr\noalign{\hrule}
&& $428  $ && $-0.02802$ && $2.54829$ & \cr\noalign{\hrule}
&& $2305 $ && $-0.01806$ && $2.55571$ & \cr\noalign{\hrule}
&& $12744$ && $-0.01240$ && $2.56472$ & \cr\noalign{\hrule}
&& $72311$ && $-0.00892$ && $2.57433$ & \cr\noalign{\hrule}
}}}
}


\vskip30pt\hbox{
{\vbox{\tabskip=0pt \offinterlineskip
\halign to 100pt{\strut#& \vrule# \tabskip=1em plus 2em&
   \hfil#& \vrule#\tabskip=0pt\cr\noalign{\hrule}

&&\omit\hidewidth Length of \hidewidth& \cr
&&\omit\hidewidth the lattice \hidewidth& \cr\noalign{\hrule}

&&  5 & \cr\noalign{\hrule}
&&  8 & \cr\noalign{\hrule}
&& 11 & \cr\noalign{\hrule}
&& 14 & \cr\noalign{\hrule}
&& 17 & \cr\noalign{\hrule}
&& 20 & \cr\noalign{\hrule}
&& 23 & \cr\noalign{\hrule}
&& 26 & \cr\noalign{\hrule}
}}}
\hskip-4.4pt
{\vbox{\tabskip=0pt \offinterlineskip
\halign to 300pt{\strut#& \vrule# \tabskip=1em plus 2em&
   \hfil#& \vrule#&
   \hfil#& \vrule#&
   \hfil#& \vrule# \tabskip=0pt\cr\noalign{\hrule}

&&\hfil&&\hfil&&\hfil& \cr
&&\omit\hidewidth Matrix Size \hidewidth&&
  \omit\hidewidth $\lambda_{\hbox{\tenrm min}}$ \hidewidth&&
  \omit\hidewidth $z_{\scriptscriptstyle L}$ \hidewidth & \cr\noalign{\hrule}

&& $1 $ && $ -1.00000$ && $ $ & \cr\noalign{\hrule} && $4 $ && $
-0.28476$ && $2.67255$ & \cr\noalign{\hrule} && $21 $ && $ -0.11943$
&& $2.72855$ & \cr\noalign{\hrule} && $103 $ && $ -0.06215$ &&
$2.70857$ & \cr\noalign{\hrule} && $553 $ && $ -0.03678$ && $2.70122$
& \cr\noalign{\hrule} && $3014 $ && $ -0.02372$ && $2.69857$ &
\cr\noalign{\hrule} && $16985$ && $ -0.01627$ && $2.69933$ &
\cr\noalign{\hrule} && $97419$ && $ -0.01168$ && $2.70193$ &
\cr\noalign{\hrule} }}} }

It is clear from an inspection of these tables that while the
convergence in each sector is reasonably good, there is a large
difference between them if we compare them between different sectors.
To see if this can be due to the presence of correction to the asymptotic
scaling form, we have tried to incorporate various forms for the correction
to the scaling. But none of these fit the data well, and at the same
time decrease the discrepancy in $z$ between different sectors. This
can be seen from the fact that the effective values of
$z_{\scriptscriptstyle L}$ do not show a significant tendency to converge
to a single value as $L$ increases, for the largest sizes reached in
our study.

One of the possible explanations for this is that  different
sectors have different gap exponents.  Though this is quite
intriguing, it is somewhat unlikely. The possible reason behind this
could be the existence of an infinite number of conserved quantities in
the model.  It is hoped that further studies will clarify this point.

However, from our data it can be concluded that the gap exponent for
all these sectors fall with in the range $z=2.55\pm 0.15$.  To get a
more precise estimate for $z$ one needs further study either of larger
size lattices, or by Monte Carlo simulations, or analytical methods.

In figure 2 we have shown a plot of $\lambda$ versus $k$ (dispersion
curve) for three different lattice sizes.  This is related to the
spectrum of the excitations of the quantum Hamiltonian $\hat W$. It is
seen that the spectrum for different sizes is qualitatively similar,
but shows a complicated, yet un-understood structure as a function of
$k$. We have also studied the same model for the case of unequal
deposition-evaporation rates (in this case there is no flip
symmetry). The range of estimated value of $z$ is the same as that for
equal deposition-evaporation rates.

The stochastic evolution of the trimer model can be mapped to the
stochastic dynamics of a string, both ends fixed to the same point, by
defining a matrix variable $U_i$ at each site [\ddmba]. This matrix
$U_i$ has information about the length of the irreducible string
corresponding to the substring from site $1$ of the lattice upto site
$i$.  Under the dynamics the length of this irreducible string
changes, and is related to the change in the matrix variables $U_i$.
Thus this model corresponds to a generalisation of the KPZ model where
the scalar height variables are replaced by matrix variables.  It is
well known that $z=1.5$ for the KPZ model [\kpz]. Our results show
that this model falls under a new universality class. It is also
different from the model studied recently by Doherty {\it et al.}
which is also a generalisation of the KPZ equation to $n$ component
variables. In their model, the dynamical exponent $z=3/2$ in one
dimension, independent of $n$, though in higher dimensions it depends
on $n$ [\doh].
\vskip .5cm
\refout
\endpage
\centerline {\bf FIGURE CAPTIONS}

\noindent 1. A plot of the effective exponent $z_{\s L}$ versus $1/L$,
where $L$ is the length of the lattice.

\noindent 2. Dispersion curve of the quantum Hamiltonian corresponding
to the trimer model [Equation {\zeroc}].

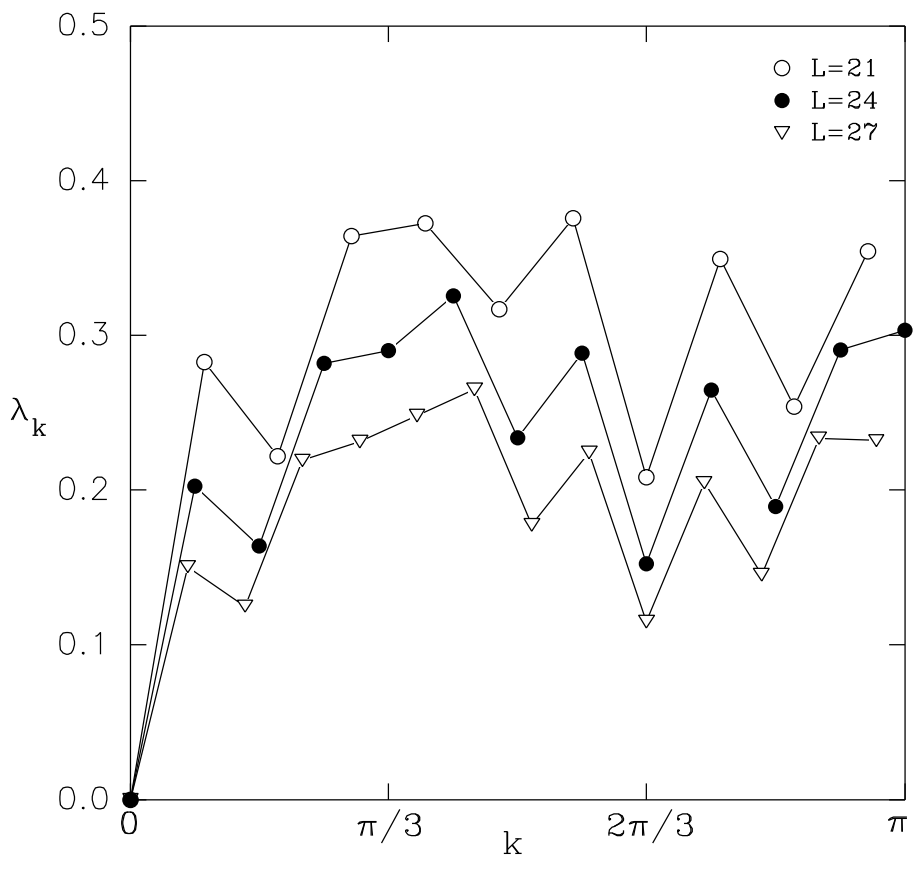
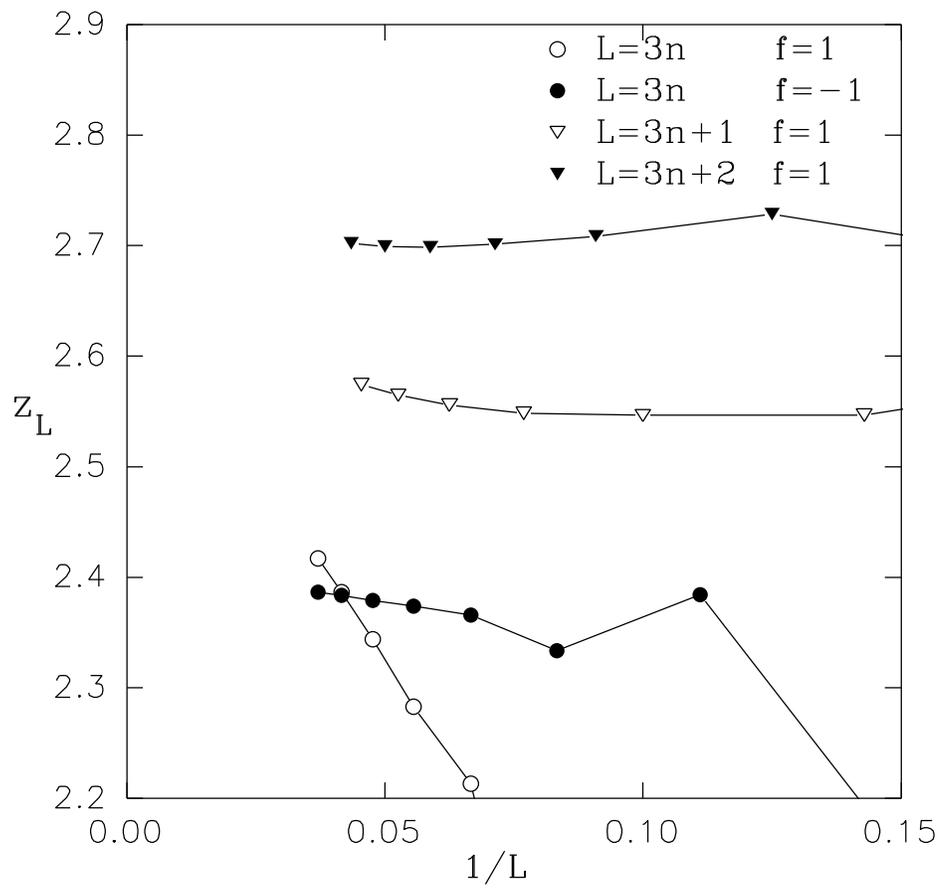
\end